\documentclass[11pt]{article}
\usepackage{float}

\usepackage{acl}

\usepackage{times}
\usepackage{latexsym}

\usepackage[T1]{fontenc}

\usepackage[utf8]{inputenc}
\usepackage{hyperref}

\usepackage{microtype}

\usepackage{inconsolata}

\usepackage{graphicx}
\usepackage{orcidlink}
\usepackage{multirow}
\usepackage{tcolorbox}
\usepackage{colortbl}
\usepackage{xcolor}

\definecolor{main}{HTML}{5989cf}    
\definecolor{darkgray}{HTML}{434343}
\definecolor{cloudgray}{HTML}{626567}
\definecolor{asphalt}{HTML}{5D6D7E}

\newcommand\tdd{Test-Driven Development}

\newtcolorbox{boxE}{
    colframe = main, 
    boxrule = 0pt, 
    toprule = 4pt 
}

\newcommand\RQONE{\textbf{RQ1}:~\textit{How far can performance improve without fine-tuning or external data augmentation?}}
\newcommand\RQTWO{\textbf{RQ2}:~\textit{Can smaller models approach larger model performance?}}
\newcommand\RQTHREE{\textbf{RQ3}:~\textit{What approach is most effective in improving vanilla (baseline) LLM accuracy?}}
\newcommand\RQFOUR{\textbf{RQ4}:~\textit{To what extent do these approaches reduce compilation errors?}}

\title{Barrier Breakers at BLP-2025 Task 2: Enhancing LLM Code Generation Capabilities through~\tdd{} and Code Interpreter}

\author{Sajed Jalil \orcidlink{0000-0003-1249-2113}\\
  Apollo Applications Group \\
  Reston, VA, USA \\
  \texttt{sajed@apolloapps.ai} \\\And
  Shuvo Saha \orcidlink{0009-0008-9267-4200} \\
  University of Dhaka \\
  Dhaka, Bangladesh \\
  \texttt{bsse0705@iit.du.ac.bd} \\\And
  Hossain Mohammad Seym \orcidlink{0009-0005-8453-3975}\\
  Islamic University of Technology\\
  Dhaka, Bangladesh\\
  \texttt{hossainseym@iut-dhaka.edu} \\}

\begin{document}
\maketitle
\begin{abstract}
Over the past few years, improving LLM code generation capabilities has been a key focus in NLP research. Despite Bengali having 242 million native speakers worldwide, it receives little attention when it comes to training LLMs. More recently, various fine-tuning and augmented generation techniques have been employed to significantly enhance code generation performance. However, they require considerable expertise and resources to utilize effectively as an end user. The goal of our work is to democratize access to powerful code generation tools in resource-constrained emerging markets, enabling users to leverage them in their native language. 

We introduce a novel approach that combines~\tdd{} (TDD) and Code Interpreter (CI), utilizing open-weight models, which improves the baseline accuracy for code generation with Bengali prompts and achieves an overall accuracy of \textbf{85\%}. Our approach requires no finetuning and proves that even the smallest models in the same family can attain up to \textbf{98\%} accuracy compared to the largest models. All of our results~\footnote{\url{https://github.com/sajedjalil/BLP25-Task-2/}} are publicly shared in GitHub for validation and reproducibility.

\end{abstract}

\section{Introduction}
Large Language Models (LLMs) have gained significant attention across various research communities since the release of ChatGPT in 2022~\footnote{\url{https://chatgpt.com/}}. Initially known as generalized text completion models, LLMs quickly found their way into more specialized tasks such as code, image, and audio generation. Specifically, the impact is visible in the code generation domain. There has been a significant transformation in the daily workflow of the software engineers with these models~\citep{jalil2023transformative}.

Despite being the 5\textsuperscript{th} most spoken language worldwide, Bengali is not included in most of the top models as a primary language for training data~\citep{raihan2025tigercoder}. Even in cross-lingual settings, most models tend to reflect Western perspectives~\citep{myung2024blend}. Additionally, prior studies have demonstrated that multilingual tokenizers are often inefficient and require additional resources during training~\citep{ali2024tokenizer}.

With these constraints in mind, we propose our work on enhancing existing open-weight LLMs of various sizes by combining~\tdd{} (TDD) and Code Interpreter (CI) without the need for fine-tuning. In this shared task with Bengali prompts, we investigate the following research questions that are crucial for advancing the field of multilingual code generation -

\RQONE

\RQTWO

\RQTHREE

\RQFOUR

\begin{figure*}
  \includegraphics[width=1\linewidth]{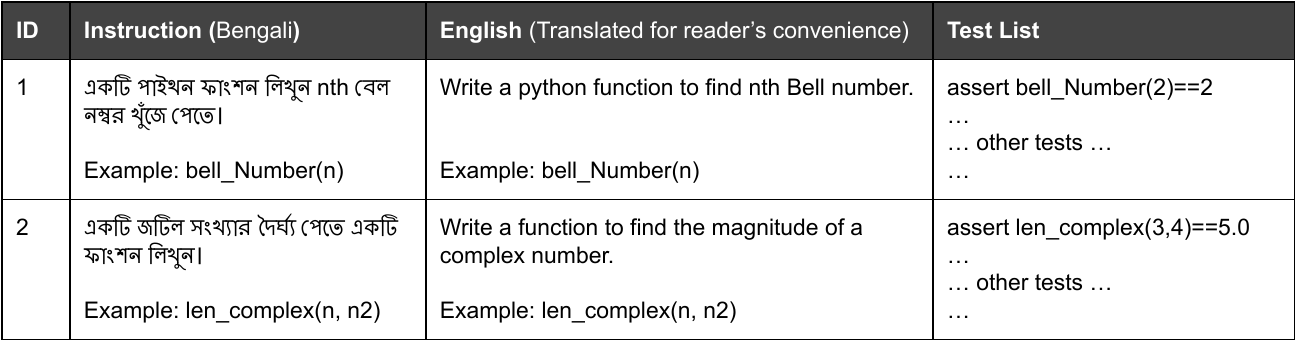}
  \caption {\label{fig:dataset}
  Example of dataset rows used in our study (English instruction is added here for readers' convenience.)}
\end{figure*}

\section{Background}
Although there have been significant prior studies in NLG and benchmarks for Bengali~\citep{bhattacharjee2022banglanlg, ekram2022banglarqa, raihan-etal-2025-mhumaneval}, the number of code generation studies using LLM is quite negligible. The only substantial study we could find is a family of finetuned models named \textit{TigerCoder}, which was evaluated for its machine translation capabilities~\citep{raihan2025tigercoder}. These findings underscore the need for further exploration using alternative techniques to improve code generation capabilities.

\tdd{} (TDD) has been a widely researched methodology in the agile software engineering domain~\citep{shull2010we, rafique2012effects}. It is the practice of writing unit tests before starting implementation to ensure software verification. This methodology has been proven to reduce code defects~\citep {williams2003test}. To the best of our knowledge, no other prior studies have explored the effects of TDD in code generation with Bengali prompts. 

Code Interpreter (CI) can act as an external tool to help LLM improve itself as a coding agent~\citep{wang2024executable}. Humans interact with LLM multiple times if the desired output is not reached~\citep{lin2025seeking}. This inspired us to utilize CI to enhance accuracy and minimize compilation errors in our study. Additionally, we employed a combined approach that incorporates TDD and CI to improve accuracy further and reduce compilation errors.

\section{Task Dataset}
The primary aim of this task was to generate Python code from Bengali instructions using LLM~\citep{raihan-etal-2025-mhumaneval, raihan-etal-2025-blp, raihan2025tigercoder}. All of our code and experimental results are publicly available on GitHub.~\footnote{\url{https://github.com/sajedjalil/BLP25-Task-2/}}

In~\autoref{fig:dataset}, a sample of the dataset is shown. The test cases evaluate the generated code. Only one test case was publicly available during the competition. The rest were hidden and could only be accessed after the submission phase had ended.


\begin{table}[ht]
    \centering
    \begin{tabular}{|l|l|}
        \hline
        \rowcolor{darkgray}
        \textcolor{white}{Model family} & \textcolor{white}{Used variants}\\
        \hline
        \hline
         Meta Llama 3.2 & 3B, 11B, 90B \\ \hline
         Meta Llama 4 & Scout 17B, Maverick 17B\\ \hline
         OpenAI gpt-oss & 20B, 120B\\ \hline
    \end{tabular}
    \caption{Distribution of LLM models and variants in our experiment.}
    \label{tab:models}
\end{table}

\section{Experiments}

Our LLM responses were generated with the AWS Bedrock platform~\footnote{\url{https://aws.amazon.com/bedrock/}}. Therefore, our selection of various models was dependent upon the availability in Bedrock. Specifically, we experimented with the following models in~\autoref{tab:models}. 

Our initiative focused on improving the accuracy of generalized LLM code generation without fine-tuning. To achieve this, we have experimented with the following five approaches -

\subsection{Vanilla (Baseline) Model}
To establish baseline accuracy with the Bengali instruction, we conducted this experiment with plain (vanilla) LLM API to determine how different LLMs perform.

\begin{figure}[htbp]
  \includegraphics[width=1\linewidth]{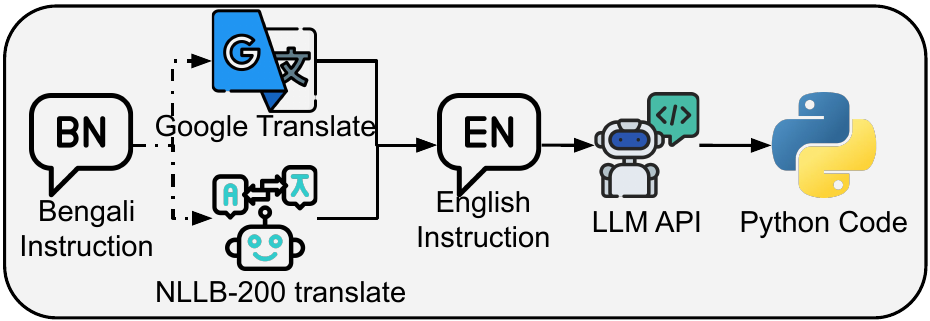}
  \caption {\label{translation}
  Two variants of Bengali to English machine translation.}
\end{figure}

\subsection{Bengali to English Machine Translation}
Since the primary language for most LLMs is English, our initial intuition was to translate the Bengali instructions into English. For this experiment, we have used two different translators - Google Translate~\footnote{\url{https://translate.google.com/}} \& NLLB-200~\citep{costa2022no}. The overall workflow of this approach is displayed in~\autoref{translation}.

\begin{figure}[t]
  \includegraphics[width=1\linewidth]{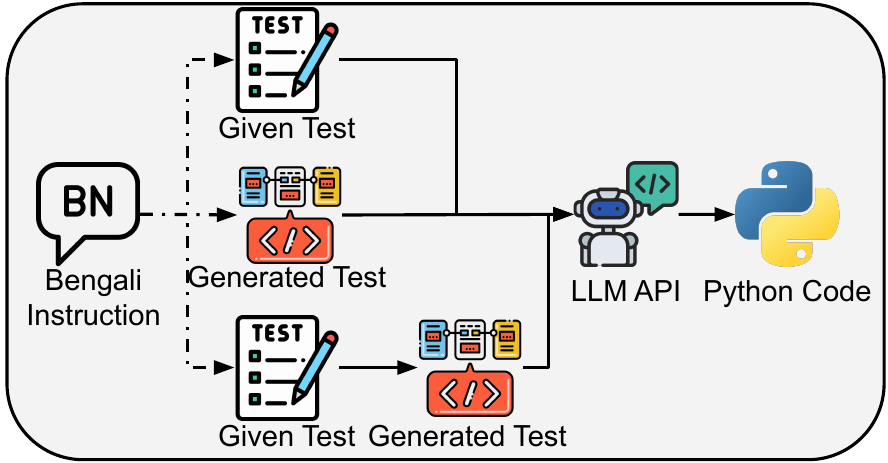}
  \caption {\label{tdd} Variants of~\tdd{} (TDD) approaches in our experiments.}
\end{figure}

\subsection{\tdd{} (TDD)}
    We experimented with three different variations of TDD in this experiment. The detailed diagram is shown in~\autoref{tdd}.
    
    \begin{enumerate}
    \item \textbf{Generated Tests} - We started with an API call to an LLM to generate up to five test cases from the given prompt. We then input these test cases, along with the given prompt, to generate our final response.

    \item \textbf{Given Test} - We injected only the publicly available assert statement (test case) from the dataset into the LLM prompt during code generation.
    \vspace{-5pt}
    
    \item \textbf{Combined} - This approach combined the above two methods. Here, we used the given test case from the dataset, along with five more LLM-generated test cases. And then, all of these test cases were used in LLM for response generation.
    \end{enumerate}

 \begin{figure} [h]
  \includegraphics[width=1\linewidth]{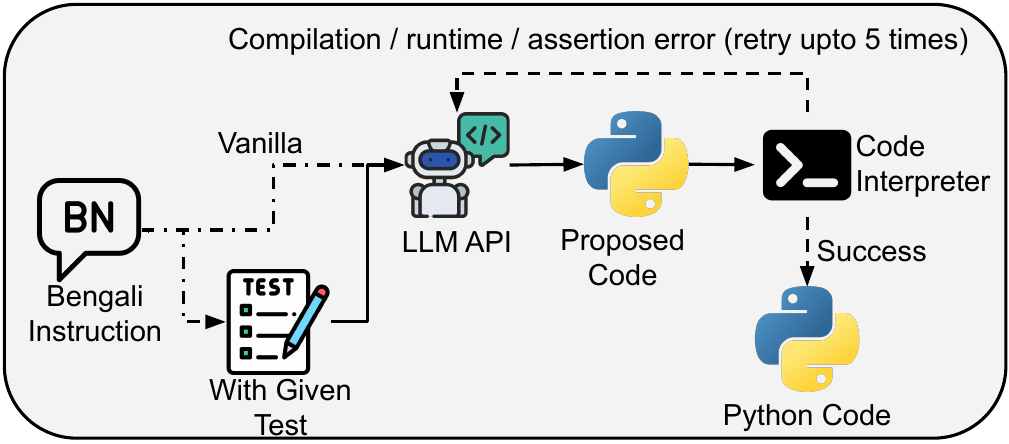}
  \caption {\label{interpreter}
  Code Interpreter with~\tdd{}~(TDD) approach.}
\end{figure}

\subsection{Code Interpreter (CI)}
    We drew inspiration for this method from how developers interact with LLMs in real life. Developers generate code from LLM and then test the code in their respective IDE or environment. If any problem is encountered, they continue the chat and share error messages with the LLM until the desired output is achieved.

    We utilized AWS Code Interpreter as a simulated Python  environment~\footnote{\url{https://docs.aws.amazon.com/bedrock/latest/userguide/agents-code-interpretation.html}}. Given a Python code, it compiled and executed it. For any errors, a detailed error message was obtained. We also set a retry limit of five to fix the generated Python code that did not compile. The error message received from the earlier execution was used as additional input for subsequent code generation. A workflow model is displayed in~\autoref{interpreter} with the vanilla path.
    
\subsection{CI+TDD}
We combined the TDD approach with the dataset provided single test case with the CI. This test case was also executed in the interpreter to verify its success. The mechanism is demonstrated in~\autoref{interpreter} with the given test path.

\begin{figure}[t]
  \includegraphics[width=1\linewidth]{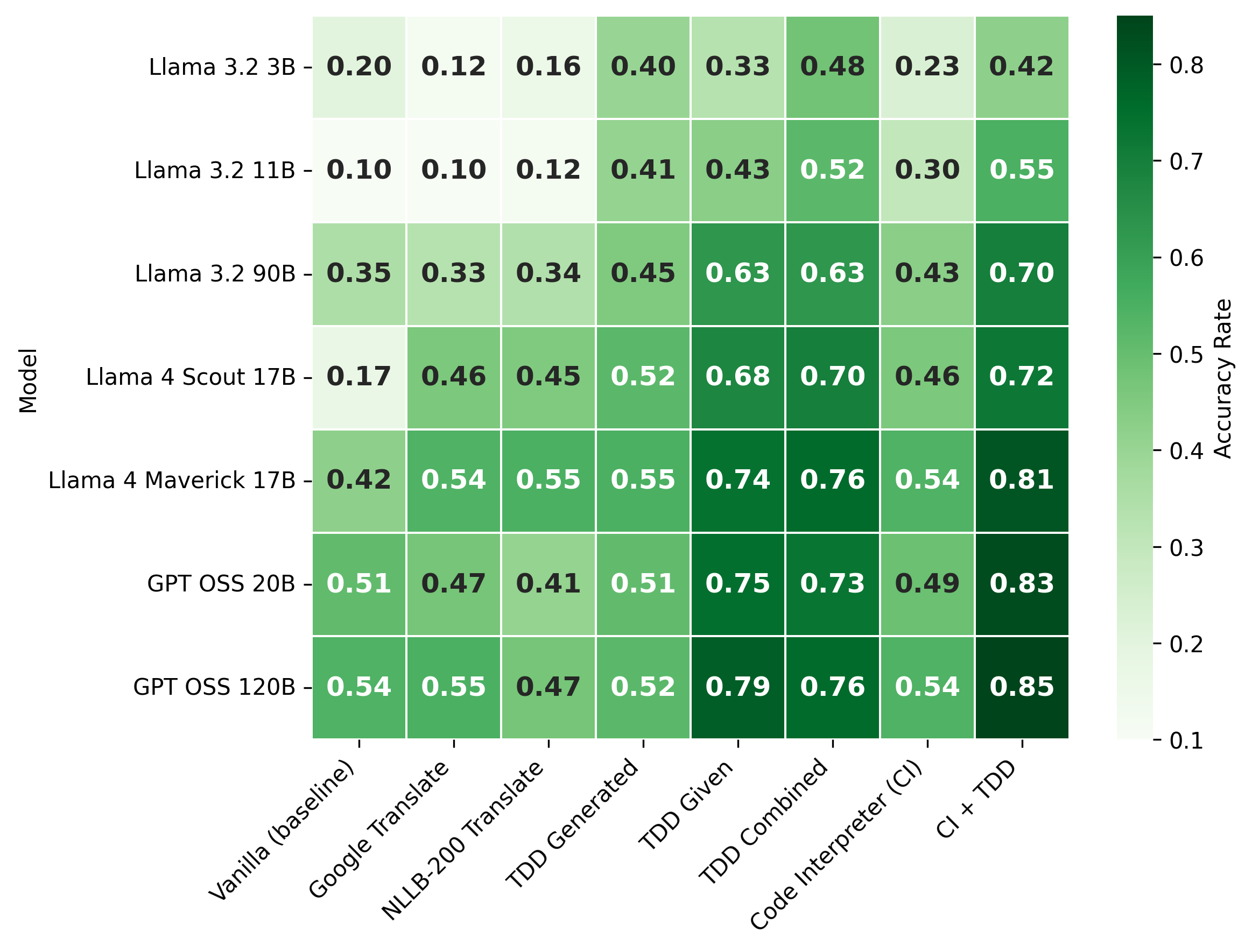}
  \caption {\label{fig:accuracy_heatmap}
  Overall accuracy heatmap of models in different approaches.}
\end{figure}

\begin{table*} [ht]
  \centering
  \begin{tabular}{|l||r|r|r|r|r|r|r|r|}
    \arrayrulecolor{white}
    \hline
    \rowcolor{darkgray}
    \textcolor{white}{Model} & \textcolor{white}{Vanilla} & \multicolumn{2}{|c|}{\textcolor{white}{Translated}} & \multicolumn{3}{|c|}{\textcolor{white}{\tdd{}}} & \multicolumn{2}{|c|}{\textcolor{white}{Code Interpreter}} \\
    \cline{3-9}
    \rowcolor{darkgray}
     && \textcolor{white}{Google} & \textcolor{white}{NLLB} & \textcolor{white}{Generated} & \textcolor{white}{Given} & \textcolor{white}{Combined} & \textcolor{white}{Vanilla} & \textcolor{white}{Given Test} \\
    \hline
    \hline
    \arrayrulecolor{black}
    
    \rowcolor{asphalt}
    \multicolumn{9}{|l|}{\textcolor{white}{Llama 3.2 models}} \\
    \hline
    3B & 19.6 & 12.4 &	16.0 &	39.6 &	33.4 &	\textbf{48.2} &	22.6 &	42.2 \\
    11B & 9.8 &	9.6 &	12.4 &	40.6 &	42.8 &	51.8 &	30.4 &	\textbf{54.8} \\
    90B & 35.0 &	32.8 &	34.2 &	44.8 &	62.8 &	63.4 &	42.6 &	\textbf{69.6} \\
    \hline
    \hline
    \rowcolor{asphalt}
    \multicolumn{9}{|l|}{\textcolor{white}{Llama 4 models}} \\
    \hline
    Scout& 16.8 &	45.8 &	45.4 &	51.6 &	67.6 &	69.6 &	45.8 &	\textbf{72.0} \\
    
    Maverick & 42.0 &	54.0 &	54.6 &	54.6 &	74.4 &	76.4 &	54.4 &	\textbf{80.6} \\
    \hline
    \hline
    \rowcolor{asphalt}
    \multicolumn{9}{|l|}{\textcolor{white}{GPT-OSS models}} \\
    \hline
    20B & 51.0 &	47.0 &	41.4 &	50.8 &	75.4 &	72.6 &	48.6 &	\textbf{82.8} \\
    120B & 54.4 & 54.6 &	46.8 &	52.2 &	79.0 &	75.6 &	54.0 &	\textbf{85.0} \\
    \hline
    
  \end{tabular}
  \caption{\label{tab:accuracy}
    Overall accuracy (\%) on varying model family and parameter size over different approaches.
  }
\end{table*}

\begin{table*} [ht]
  \centering
  \begin{tabular}{|l||r|r|r|r|r|r|r|r|}
    \arrayrulecolor{white}
    \hline
    \rowcolor{darkgray}
    \textcolor{white}{Model} & \textcolor{white}{Vanilla} & \multicolumn{2}{|c|}{\textcolor{white}{Translated}} & \multicolumn{3}{|c|}{\textcolor{white}{\tdd{}}} & \multicolumn{2}{|c|}{\textcolor{white}{Code Interpreter}} \\
    \cline{3-9}
    \rowcolor{darkgray}
     && \textcolor{white}{Google} & \textcolor{white}{NLLB} & \textcolor{white}{Generated} & \textcolor{white}{Given} & \textcolor{white}{Combined} & \textcolor{white}{Vanilla} & \textcolor{white}{Given Test} \\
    \hline
    \hline
    \arrayrulecolor{black}
    
    \rowcolor{asphalt}
    \multicolumn{9}{|l|}{\textcolor{white}{Llama 3.2 models}} \\
    \hline
    3B & 21.4 &	61.8 &	38.4 &	8.4 &	7.6 &	5.8 &	\textbf{0.2} &	0.8 \\
    11B & 67.8 &	71.2 &	61.0 &	5.2 &	0.2 &	2.0 &	\textbf{0.0} &	0.2 \\
    90B & 8.8 &	15.4 &	9.4 &	\textbf{0.2} &	\textbf{0.2} &	0.4 &	0.4 &	\textbf{0.2} \\
    \hline
    \hline
    \rowcolor{asphalt}
    \multicolumn{9}{|l|}{\textcolor{white}{Llama 4 models}} \\
    \hline
    Scout & 66.4 &	1.2 &	1.4 &	3.6 &	0.8 &	0.8 &	\textbf{0.2}	 & 2.6\\
    
    Maverick & 19.2 &	0.2 &	0.2 &	1.2 &	\textbf{0.0} &	0.2 &	0.2 & \textbf{0.0} \\
    \hline
    \hline
    \rowcolor{asphalt}
    \multicolumn{9}{|l|}{\textcolor{white}{GPT-OSS models}} \\
    \hline
    20B & 1.0 &	1.8 &	1.8 &	1.8 &	1.2 &	1.2 &	\textbf{0.2} &	\textbf{0.2} \\
    120B & 0.2 &	0.4 &	0.2 &	0.4 &	0.8 &	1.2 &	\textbf{0.0}	& 0.2\\
    \hline
    
  \end{tabular}
  \caption{\label{tab:compile-error}
    Overall compilation errors occurrence (\%) on varying model family and parameter size.
  }
\end{table*}

\begin{figure*}[ht]
  \includegraphics[width=1\linewidth]{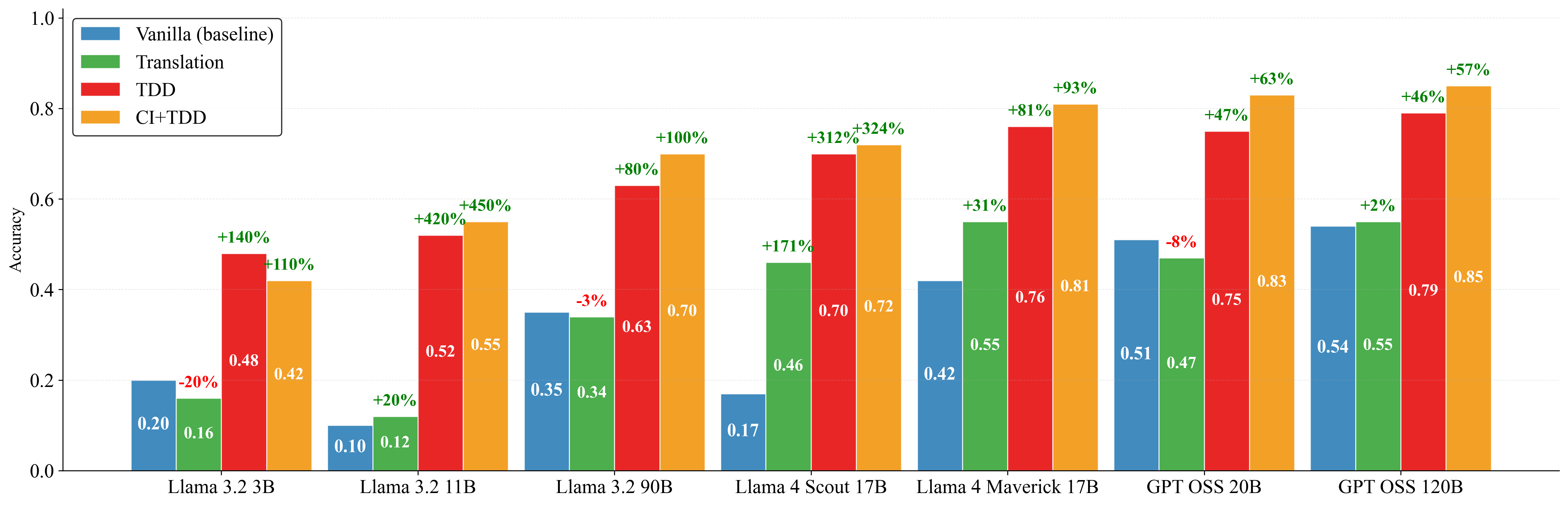}
  \caption {\label{fig:increase}
  Model accuracy comparisons of baseline vs. our approaches (with increase/decrease in percentage).}
\end{figure*}

\section{Results \& Analysis}
The results of our experiments are provided in~\autoref{tab:accuracy} and~\autoref{tab:compile-error}. We group the data by model family and model parameters. The best outcome for each model is represented by bold text.

~\textbf{\RQONE}

In our investigation, we obtained several interesting findings.~\autoref{fig:accuracy_heatmap} demonstrates the overall accuracy score on the test phase. It is distinctly evident from the heatmap that vanilla (baseline) accuracy can be improved significantly with TDD and CI. Except for the Llama 4 models, machine translation from Bengali to English did not provide significant improvement. Instead, it harmed the overall accuracy compared to the non-translation approach.

On the other hand, we observed an impressive increase in accuracy compared to the baseline in~\autoref{fig:increase}. The CI+TDD approach improved the accuracy across all models by \textbf{+57\%} to \textbf{+450\%}. The TDD approach improves the baseline by \textbf{+47\%} to \textbf{+420\%}. Bengali to English machine translation has a change factor from \textbf{-20\%} to \textbf{+171\%}.

\begin{boxE}
\textbf{\small Compared to baseline (54\%), overall accuracy can be improved up to 85\% using our proposed techniques.}
\end{boxE}

\vspace{10pt}

\textbf{\RQTWO}

In the Llama 3.2 model family, using the TDD 3B variant (48\%) exceeds the baseline accuracy of the 90B variant (35\%). Comparing the best outcome for each variant, we observed that 3B could reach \textbf{67\%} of the performance of 90B and \textbf{87\%} of the performance of 11B.

For the Llama 4 model family, Scout can exceed the Maverick baseline by 71\% using CI+TDD. When comparing best outcomes, Scout can achieve up to \textbf{89\%} of Maverick's performance.

Lastly, in the gpt-oss variants, the 20B variant using CI+TDD surpasses the 120B baseline by 54\%. In best-case scenarios for both, 20B can reach up to \textbf{98\%} of the performance of 120B. Our results further confirm the claims made by another prior study~\citep{belcak2025small}.

\begin{boxE}
\textbf{\small In the same model family, the smallest model can attain up to 98\% accuracy of the largest model.}
\end{boxE}
\vspace{10pt}

\textbf{\RQTHREE} 

As~\autoref{fig:increase} indicates that the best outcome is always from CI+TDD except for the Llama 3.2 3B model. It should be noted that both TDD and CI+TDD performed significantly better than baseline in all models.

\begin{boxE}
\textbf{\small Combination of~\tdd{} (TDD) and Code Interpreter (CI) yields the largest jump in accuracy. }
\end{boxE}

\begin{figure} [h]
  \includegraphics[width=1\linewidth]{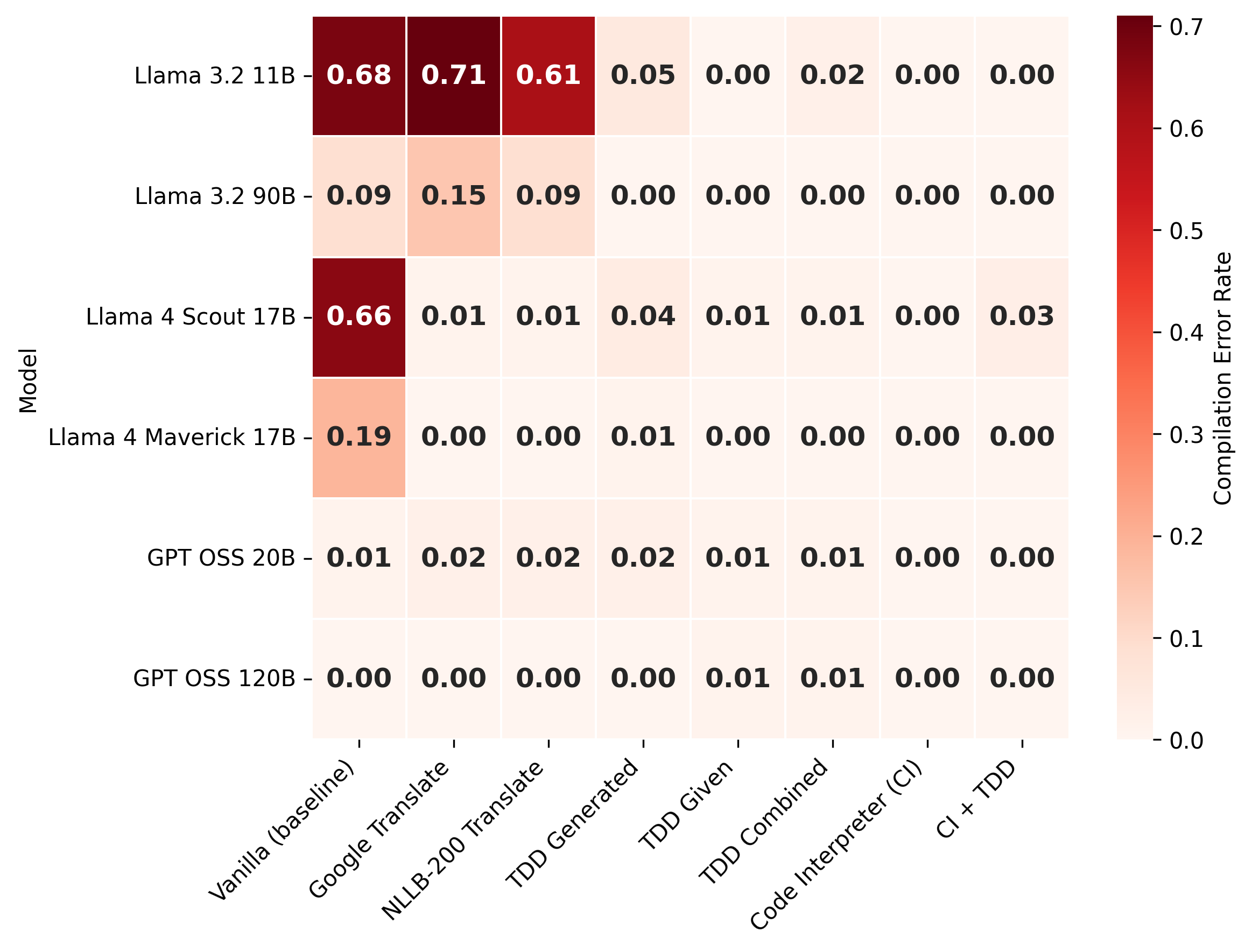}
  \caption {\label{fig:compilation_errors_heatmap}
  Compilation error rate heatmap of models in different approaches.}
\end{figure}

\textbf{\RQFOUR}

In terms of total compilation errors in the generated code, a similar trend is visible as the accuracy rate.~\autoref{fig:compilation_errors_heatmap} demonstrates both TDD and CI approaches have nearly eliminated all compilation errors with rates approaching \textbf{0\%}. Translation helped reduce compilation errors only in the Scout model families. An interesting trend is observed in the gpt-oss model family, whose baseline compilation errors are nearly zero, suggesting it may contain inherent mechanisms to address compilation issues.

\begin{boxE}
\textbf{\small Both TDD and CI reduce compilation errors, whereas Bengali to English machine translation increases the error count in most cases.}
\end{boxE}

\section*{Conclusion}
This study successfully introduced a novel approach that combines~\textbf{\tdd{} (TDD) and Code Interpreter (CI)} to improve code generation accuracy for Bengali prompts utilizing open-weight LLMs. Our findings demonstrate that this strategy yields significant improvements without requiring resource-intensive fine-tuning or the use of external data for augmentation. The \textbf{CI+TDD methodology} was the most effective, increasing overall baseline accuracy by up to {450\%} and virtually eliminating compilation errors across all models tested. Furthermore, our research suggests that using these strategies, even the smallest models in the same family can achieve up to \textbf{{98\%} accuracy} when compared to the largest models of the same family. We strongly believe the exact mechanism can be applied to other underrepresented languages, similar to Bengali, and increase access to high-performing code generation tools in resource-constrained emerging markets.

\section{Limitations}
Our study is limited in context, as we only checked a subset of open-weight models available on the AWS Bedrock API. This restricted us from using several other popular models not available on that platform, such as Qwen3 and Gemma3. Moreover, we did not explore how our approach would perform in larger and complex coding tasks, as opposed to the single method-based problems provided in the shared-task dataset.

\section*{Acknowledgments}
We greatly appreciate Apollo Applications Group~\footnote{\url{https://www.apolloapps.ai/}} for granting us access to the AWS resources necessary to perform this study. And special thanks to Marium Binte Ibrahim Ema and Jeffrey Dershewitz for their reviews.

\bibliography{custom}

\appendix



\end{document}